\documentclass[a4paper, 10 pt, conference]{ifacconf}  

\usepackage{graphicx}      
\usepackage{natbib}        

\usepackage{graphics} 
\usepackage{epsfig} 
\usepackage{amsmath, bm} 
\usepackage{amssymb}  
\usepackage{amsfonts}
\usepackage{mathtools}
\usepackage{tikz}
\usepackage{marvosym}

\usepackage{enumitem}
\usepackage{braket}
\usepackage{listings}
\usepackage{textcomp}
\definecolor{commentgreen}{rgb}{0.133, 0.543, 0.133}

\newcommand{\bs}[1]{\ensuremath{\mathbf{#1}}}

\begin{document}

\begin{frontmatter}

\title{PoCET: a Polynomial Chaos Expansion Toolbox for \textsc{Matlab}} 


\author[First]{Felix Petzke} 
\author[Second]{Ali Mesbah} 
\author[First]{Stefan Streif} 

\address[First]{Automatic Control \& System Dynamics Lab at Technische Universit\"at Chemnitz, Faculty for Electrical Engineering and Information Technology, Chemnitz, Germany (e-mail: {\{felix.petzke, stefan.streif\}@etit.tu-chemnitz.de}).}
\address[Second]{Department of Chemical and Biomolecular Engineering, University of California, Berkeley, CA, USA (e-mail: {mesbah@berkeley.edu}).}

\begin{abstract}
We introduce PoCET: a free and open-scource Polynomial Chaos Expansion Toolbox for \textsc{Matlab}, featuring the automatic generation of polynomial chaos expansion (PCE) for linear and nonlinear dynamic systems with time-invariant stochastic parameters or initial conditions, as well as several simulation tools. It offers a built-in handling of Gaussian, uniform, and beta probability density functions, projection and collocation-based calculation of PCE coefficients, and the calculation of stochastic moments from a PCE. Efficient algorithms for the calculation of the involved integrals have been designed in order to increase its applicability. 
PoCET comes with a variety of introductory and instructive examples. Throughout the paper we show how to perform a polynomial chaos expansion on a simple ordinary differential equation using PoCET, as well as how it can be used to solve the more complex task of optimal experimental design. 
\end{abstract}
\end{frontmatter}


\section{Introduction}
Uncertainty quantification and propagation have always been irrefutably relevant topics in natural sciences and engineering, since almost every real-life process is at some point subject to disturbances or parametric uncertainties, determining its outcome in a somewhat random way. Of the several approaches available to describe and propagate probability densities, this paper focusses on the one devised by Norbert Wiener in 1938: the polynomial chaos expansion (PCE). In its core, a PCE is a series expansion method for density functions, based on distribution-specific polynomial basis functions. Through this expansion, any system with time-invariant uncertainties (e.g. in initial conditions or parameters) can be transformed into a set of deterministic equations. While the transformed system allows for very fast (online) simulation times compared to sampling-based approaches, the (offline) transformation itself can be computationally very expensive, since it requires solving a large number of integrals.

However, due to the increase of computational power in recent years, PCE has become more applicable to numerical problems. Its main applications range from uncertainty quantification (\cite{Bijl06,Eldred09, Eldred12, Savin19a}), to general stochastic differential equations (\cite{Xiu02}, \cite{McKenzie12}) and systems with probabilistic uncertainties (\cite{Schick13,Braatz13}). Furthermore, approaches dealing with time-variant uncertainties have been developed (\cite{Gerritsma10}). And specifically the possibility of fast uncertainty propagation has made PCE a widely applied tool for different online optimization problems like stochastic MPC (\cite{Mesbah14,Paulson14}) or experimental design (\cite{Streif14, Mesbah15}). 

This paper presents PoCET, a \textsc{Matlab} toolbox specifically designed to perform a projection-based polynomial chaos expansion on dynamic systems with time-invariant uncertainties. Our aim was to automate the processes of choosing and setting up a sufficiently large polynomial basis, calculating the involved integrals and expanded system matrices, and writing the respective \textsc{Matlab} function files required to solve the expanded system. While there are numerous toolboxes on uncertainty quantification in \textsc{Matlab} already available (notably UQLab; \cite{uqlab}) they do not provide an automated method for uncertainty propagation in the Galerkin projection PCE framework for ODEs (cf. Sec. \ref{sec:pce}), which is particularly useful for polynomial systems (cf. Sec. \ref{sec:nonlinear-systems}).\\
PoCET is an open-source toolbox\footnote{The latest version is available for download on GitHub. See \texttt{www.tu-chemnitz.de/etit/control/research/PoCET/} for more information.} 
that comes with detailed introductory and instructive examples and it is available for free. Note that PoCET requires \textsc{Matlab}'s Symbolic Math Toolbox for parsing; apart from that it is stand-alone.

\textit{Notation:} $N_{\xi}$ and $P$ denote the number of independent uncertain variables and the order of the PCE, respectively. 
$\varphi_n$ is a polynomial of order $n$ from the polynomial basis $\bm{\Phi}$. 
States, their expansions, and the vector of their expansions are denoted by $x$, $\hat{x}_{i}$, and $\bm{\hat x}$, respectively.
$\mu$ is a probability density function with corresponding support $\Omega$ and the $m$-th stochastic moment $\nu^{(m)}$.
Gaussian variables are denoted by $\xi\sim\mathcal{N}(\nu^{(1)},\tilde\nu^{(2)})$, uniformly distributed ones by $\xi\sim\mathcal{U}(a,b)$, and beta-distributed ones by $\xi\sim\mathcal{B}(\alpha,\beta)$ or $\xi\sim\mathcal{B}_4(\alpha,\beta,l,u)$ for the standard or 4-parameter beta distributions, respectively. 
The Kronecker product is denoted by $\otimes$. We use the short-hand notation $\sum_{i_1=0}^N+\dots+\sum_{i_m=0}^N=\sum_{i_1,\dots,i_j=0}^N$ for a multi-indexed summation.


\section{Polynomial chaos expansion}\label{sec:pce}
This section very briefly introduces the main idea and mathematical foundation, and highlights common challenges and limitations. Thorough introductions can be found in \cite{McKenzie12, OHagan13, Sudret14}.

\subsection{General idea and mathematical foundation}
\vspace{-0.5em}The general (truncated) PCE of dimension $N_\xi$ and order $P$ of a random variable $v$ is given by
\begin{align}
x\approx&\sum_{i=0}^{\tilde{P}-1} \hat{x}_i\varphi_i\big(\bm{\hat \xi}\big) = \bm{\hat x}^T\bm{\Phi}\big(\bm{\hat \xi}\big)\label{eq:pce_allg}
\end{align}
with $\tilde{P}=\frac{(N_\xi+P)!}{N_\xi! P!}$, $\bm{\hat{x}},\bm{\Phi}\in\mathbb{R}^{\tilde{P}}$, and $\bm{\hat \xi}\in\mathbb{R}^{N_\xi}$. The polynomial basis $\bm{\Phi}$ consists of $P$ orthogonal polynomials $\varphi_n$ with
\begin{align}
\label{eq:scalprod}
\Braket{\varphi_i,\varphi_j}=\int_\Omega\varphi_i(\xi)\varphi_j(\xi)\rho(\xi)d\xi=\lambda_{i}\delta_{ij}
\end{align}
with $\lambda_i\in\mathbb{R}$ and $\delta_{ij} := \{1 \text{ if } i=j,\;0 \text{ else}\}$. The actual polynomials $\varphi_n$ depend on the probability densitiy function (PDF) of $\xi$ and for many common distributions such orthogonal bases are known (cf. \cite{Savin19a} and references therein). So far, PoCET supports Gaussian, uniform and beta distributions (i.e. Hermite, Legendre, and Jacobi polynomial bases, respectively).\\
Furthermore, there are two general approaches to calculate the PCE coefficients: non-intrusive (or projection based, cf. \cite{Eldred08}), henceforth called \textbf{pPCE}, and intrusive (or collocation based) methods, henceforth called \textbf{cPCE}. PoCET was specifically designed for Galerkin projection based PCE, however, collocation is still supported since the former is only applicable to polynomial systems. A detailed comparison between both methods can be found in \cite{Eldred09}.

\subsection{PCE for dynamic systems}
\vspace{-0.5em}Consider the ordinary differential equation (ODE)
\begin{align}
\dot x(t) = a(\xi)x(t)
\label{eq:simpleode}
\end{align}
where $a(\xi)$ is stochastic and time-invariant. The extended system is then given by (cf. equation \eqref{eq:pce_allg})
\begin{align}
\sum_{n=0}^{\tilde{P}-1}\dot{\hat{x}}_n\varphi_n = \sum_{j=0}^{\tilde{P}-1} \hat{a}_j\varphi_j\sum_{k=0}^{\tilde{P}-1} \hat{x}_k\varphi_k.
\end{align}
In order to solve this extended system of ODEs, we project the entire system onto $\varphi_i\in\bm{\Phi}$, which yields
\begin{align}
\sum_{n=0}^{\tilde{P}-1}\dot{\hat{x}}_n\Braket{\varphi_n,\varphi_i} = \sum_{j,k=0}^{\tilde{P}-1}\hat{a}_j\hat{x}_k\Braket{\varphi_j\varphi_k,\varphi_i},
\label{eq:orth-projection}
\end{align}
and further, due to orthogonality of $\varphi_n$ and $\varphi_i$,
\begin{align}
\dot{\hat{x}}_i = \frac{1}{\Braket{\varphi_i,\varphi_i}}\sum_{j,k=0}^{\tilde{P}-1}\hat{a}_j\hat{x}_k\Braket{\varphi_j\varphi_k,\varphi_i}=\sum_{j=0}^{\tilde{P}-1} \bm{E}^{(1)}_j \hat{a}_j\bm{\hat x}
\label{eq:pceode}
\end{align}
with 
\begin{align}
\bm{E}^{(1)}_j=
\begin{bmatrix} 
e_{j00} & e_{j10} & \cdots & e_{jS0} \\ 
e_{j01} & e_{j11} & \cdots & e_{jS1} \\ 
\vdots & \vdots & \ddots & \vdots \\
e_{j0S} & e_{j1S} & \cdots & e_{jSS}
\end{bmatrix},\;
e_{jki}:=\frac{\Braket{\varphi_j\varphi_k,\varphi_i}}{\Braket{\varphi_i,\varphi_i}},
\label{eq:coeffmat}
\end{align}
and $S=\tilde P-1$.
The superscript $(1)$ denotes the order of the coefficient matrix $E$, which is equal to the sum of orders of all states that appear in the respective monomial (cf. Sec. \ref{sec:nonlinear-systems} for more details). In order to solve \eqref{eq:pceode}, we calculate the values of the coefficients $\hat{a}_j$ as
\begin{align}
\label{eq:calc_aj}
\hat{a}_j=\frac{\Braket{a(\xi),\varphi_j}}{\Braket{\varphi_j,\varphi_j}}.
\end{align}

Alternatively, we can calculate these coefficients with a least-squares optimization approach using samples of the respective random variables as
\begin{align}
\label{eq:calc_aj_coll}
\hat{\bs{a}}^\star=\arg \underset{\hat{\bs{a}}}{\min} \left(\bs{y}-\bm{\Phi}\bs{\hat{a}}\right)^\top\left(\bs{y}-\bm{\Phi}\bs{\hat{a}}\right),
\end{align}
where $\bs{y}=[a(\xi_1),\dots,a(\xi_q)]$ is the vector of observations and $\xi_1,\dots,\xi_q$ are samples of the random variable $\xi$.
This approach is generally referred to as stochastic regression or collocation. 

\subsection{Challenges and limitations}
\vspace{-0.5em}There are several common challenges regarding the implementation of a PCE. Setting up the extended system \eqref{eq:pceode} requires solving $\tilde{P}^{p+2}$ integrals, where $p$ is the highest order of monomials. In PoCET these integrals are solved using adaptive quadrature rules, which are well known for all of the supported polynomials (cf. \cite{Press92}). 
Furthermore, the $n$-th stochastic moment from the PCE coefficients after solving the extended system can be calculated as
\begin{align}
\begin{split}
\nu^{(m)}(x)&=\int_\Omega x^m(\xi)\mu(d\xi)\\
&\approx\sum_{i_1,\dots,i_m=0}^{\tilde{P}-1}\hat{x}_{i_1}\cdots \hat{x}_{i_m}\Braket{\varphi_{i_1}\cdots,\varphi_{i_m}}\\
&=\big(\underbrace{\bm{\hat x}\otimes\cdots\otimes\bm{\hat x}}_{m \text{ times}}\big)^T\bm{\hat{E}}^{(m)}_\nu
\end{split}\label{eq:moments}
\end{align}
with $\bm{\hat{E}}^{(m)}_\nu=\big[\epsilon_{0\dots0},\;\dots,\;\epsilon_{0\dots S},\;\dots,\;\epsilon_{1\dots S},\;\dots,\;\epsilon_{S\dots S}
\big]^\top \in \mathbb{R}^{\tilde{P}^m}$ and $\epsilon_{i_1\dots i_m}:=\Braket{\varphi_{i_1}\cdots\varphi_{i_{m-1}},\varphi_{i_m}}$
which requires solving another $\tilde{P}^{m}$ integrals. 
Lastly, modeling time-varying random signals (e.g. noise) still remains challenging and has only been approached in recent years (see, e.g., \cite{Paulson19}). 


\section{Main Functionalities of P{\small o}CET}\label{sec:pocet}
Our aim of designing PoCET was to facilitate the application of a polynomial chaos expansion to dynamic systems in \textsc{Matlab}. In order to demonstrate its usage, we consider the autonomous system
\begin{align}
\begin{split}
\dot x(t) &= -a(\xi)x(t),\;x(0) = 2,\\
a&\sim\mathcal{\mathcal{B}}(2,2),
\end{split}
\label{eq:exampleode}
\end{align}
where $a$ is beta-distributed on the interval $[0,1]$ with shape parameters $\alpha=\beta=2$. The following sections provide a step-by-step demonstration of how to define, expand, and simulate this system using PoCET.

\subsection{System definition}\label{sec:sysdef}
\vspace{-0.5em}There are four pre-defined building blocks we can use to define systems in PoCET: states, parameters, inputs, and outputs, which are each defined as structures (\textsc{Matlab} stucts). States are time variant variables that are defined by an ODE or difference equation, respectively, as well as a possibly uncertain initial condition. Parameters are time invariant variables and therefore only defined by a probability distribution. Inputs are possibly time variant functions explicitly defined by the user. And outputs are time-invariant functions of the states. 

Defining system \eqref{eq:exampleode} amounts to the code
\begin{lstlisting}
states(1).name = 'x'; % variable name
states(1).pdf  = 'dirac'; % variable's probability density function 
states(1).data = 2; % distribution's parameters
states(1).rhs  = '-a*x'; % right hand side of ODE

parameters(1).name = 'a';
parameters(1).pdf  = 'beta'; 
parameters(1).data = [2, 2]; 
\end{lstlisting}\vspace{-0.6em}
where the expected input for the property \texttt{data} depends on the chosen distribution (e.g. a uniform distribution is defined by its lower and upper bound; cf. \texttt{help \textcolor{blue}{PoCET}}). 



\subsection{Generating the expanded system}\label{sec:sysgen}
\vspace{-0.5em}This step includes one of the main contributions of PoCET: the automatic generation of the extended system \eqref{eq:pceode} from a specified ODE and writing the respective function files required for a simulation. First we define the desired order $P$ of the PCE and then call the following functions using the structures defined in the previous step.
\begin{lstlisting}
pce_order = 3; % desired order of the PCE
pce_sys = PoCETcompose(states,parameters,[],[],pce_order); % analyze input and generate PCE
PoCETwriteFiles(pce_sys,'PCEODE.m',[],'NOMODE.m') % create .m-files for ODE functions
\end{lstlisting}\vspace{-0.6em}
The main work here is done by the function \textcolor{blue}{\texttt{PoCETcompose}}, which checks the input structures for discrepancies, completeness, and applicability of projection based PCE. If the latter is true, it also computes all of the involved PCE coefficients $\bm{E}_j^{(p)}$ (cf. Eq. \eqref{eq:coeffmat}) via Galerkin projection. The structure \texttt{pce\_sys} contains these coefficients, as well as the initial conditions for the extended system. \texttt{\textcolor{blue}{PoCETwriteFiles}} creates .m-function files in the current directory which contain the expanded ODE system (\texttt{PCEODE.m}) and the nominal ODE system (\texttt{NOMODE.m}), respectively, the former of which is used for pPCE simulations and the latter for cPCE or Monte Carlo simulations. 

\subsection{System simulation}
\vspace{-0.5em}We carry out the simulation of the expanded system using the system structure and the automatically generated ODE function files from Step \ref{sec:sysgen} by employing either the function \textcolor{blue}{\texttt{PoCETsimGalerkin}} or \textcolor{blue}{\texttt{PoCETsimCollocation}} for pPCE or cPCE, respectively. Both functions output the solution of the PCE system, i.e. the values of the extended states $\hat{\bs{x}}_{i}(t)$ over the specified time horizon. Additionally, the function \textcolor{blue}{\texttt{PoCETsimMonteCarlo}} provides a quick way to perform a Monte Carlo simulation of the specified system. 
Note that all simulation routines offered by PoCET are essentially wrappers for \textsc{Matlab}'s built-in ODE solvers and therefore all of the usual options can be specified. To do so, just create a structure including simulation time, step size, the desired solver, and its respective options.
\begin{lstlisting}
simoptions.tspan = [0 1]; % simulation time span
simoptions.dt = 0.005; % output time step size
simoptions.solver = 'ode15s'; % desired solver
simoptions.setup = odeset; % solver options
\end{lstlisting}

\subsubsection{Projection-based PCE}
\vspace{-0.5em}Using a Galerkin projection approach generally allows for very fast online computation times, since we already calculated all PCE coefficients offline in the previous step. We solve the expanded system by calling the function \textcolor{blue}{\texttt{PoCETsimGalerkin}}. 
\begin{lstlisting}
gal_results = PoCETsimGalerkin(pce_sys,'PCEODE',[],simoptions); % system simulation
\end{lstlisting}\vspace{-0.6em}
The structure \texttt{gal\_results} contains fields for all states and outputs of the system which in turn contain the solution of their respective expansions over time. These can then be used for either quick sampling or to calculate the moments of the original variables. 
We can also solve the system manually using the generated ODE function files and the initial conditions from the system structure \texttt{pce\_sys}, which generally yields  faster computation times.

\subsubsection{Collocation-based PCE and sampling}
Collocation methods provide an alternative way to calculate the PCE coefficients 
using samples of the involved random variables (cf. Eq. \eqref{eq:calc_aj_coll}). PoCET has a dedicated routine called \texttt{\textcolor{blue}{PoCETsample}} for drawing samples as they are used for several different purposes. As mentioned above, PoCET's cPCE routine \textcolor{blue}{\texttt{PoCETsimCollocation}} uses the nominal system ODE for simulation as well as samples from the stochastic basis $\xi$.
\begin{lstlisting}
n_samples = 10000; % number of samples
basis = PoCETsample(pce_sys,'basis',n_smpls); % draw samples for stochastic basis 
col_results = PoCETsimCollocation(pce_sys,'NOMODE', [],basis,simoptions); % simulate system
\end{lstlisting}\vspace{-0.6em}
The structure \texttt{col\_results} again contains the solutions of the expanded variables.

\subsubsection{Monte Carlo simulation}
Similar to cPCE, Monte Carlo simulations use the nominal ODE as well as samples on the involved random variables. In this case, however, we do not use samples of the stochastic basis $\xi$ but samples of the actual random variable $a(\xi)$, which is done by calling \texttt{\textcolor{blue}{PoCETsample}} with the option \texttt{variables}. 
\begin{lstlisting}
mc_samples = 10000; % number of samples
vars = PoCETsample(pce_sys,'variables',mc_samples); % draw samples for system variables
mc_results = PoCETsimMonteCarlo(pce_sys,'NOMODE', [],vars,simoptions); % system simulation
\end{lstlisting}\vspace{-0.6em}
The resulting structure \texttt{mc\_results} contains the sampled trajectories of the original states. 
\subsection{Moment calculation and PDF fitting} 
\begin{figure}[b]
\includegraphics[width=\columnwidth]{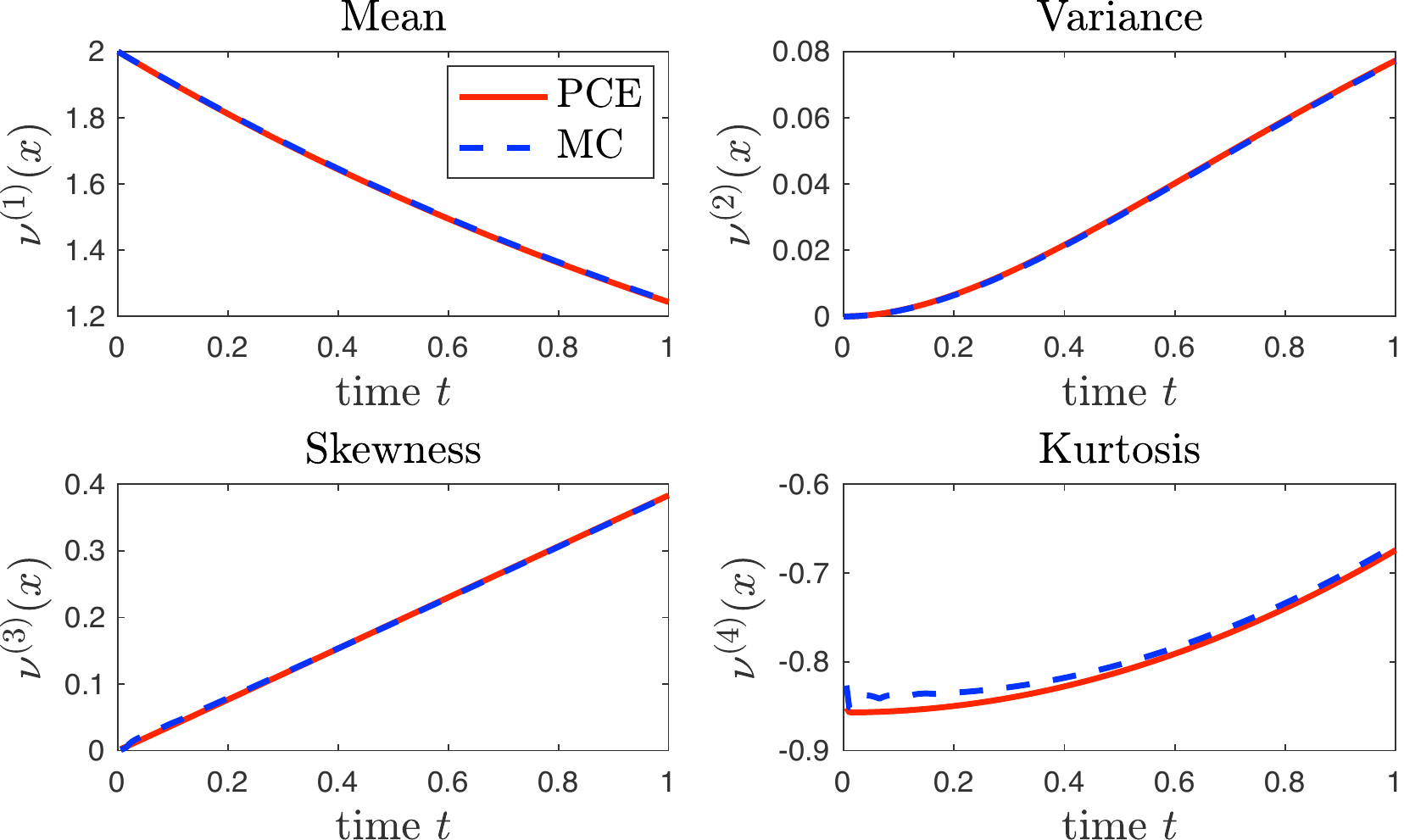}
\caption{Comparison of trajectories for the first four moments of of state $x$ in \eqref{eq:exampleode} obtained by a pPCE of order 3 and a Monte-Carlo (MC) simulation using $10^6$ samples. Simulation and moment calculation took 0.08\,s for the pPCE and 333.29\,s for the MC approach.}
\label{fig:ex-simple}
\end{figure}

\vspace{-0.5em}After solving the expanded system we now show how to retrieve stochastic information about the original states $x(\xi)$. In order to compute their first $m$ moments $\nu^{(m)}(x)$, we first have to calculate the coefficients $\bm{\hat{E}}^{(m)}_\nu$ (cf. Eq. \eqref{eq:moments}), which is done via the function \textcolor{blue}{\texttt{PoCETmomentCoeffs}}. Afterwards we call \textcolor{blue}{\texttt{PoCETcalcMoments}} to calculate the actual moments.
\begin{lstlisting}
m = 4; % highest order to be calculated
MomCoeffs = PoCETmomentCoeffs(pce_sys,m); % calculate coefficients for moment calculation
results.x.moments = PoCETcalcMoments(pce_sys,MomCoeffs,results.x.pcvals); % calc. moments
\end{lstlisting}\vspace{-0.6em}
The resulting field \texttt{moments} is an $m\times k$ matrix where $k$ is the number of time steps in the simulation. 
Fig. \ref{fig:ex-simple} shows the resulting trajectories for the the first four moments in comparison to a Monte-Carlo simulation, which can be done as described above. For an analysis of the convergence properties of general polynomial chaos the interested reader is referred to \cite{Ernst12}.

\begin{figure}[t]
\includegraphics[width=\columnwidth]{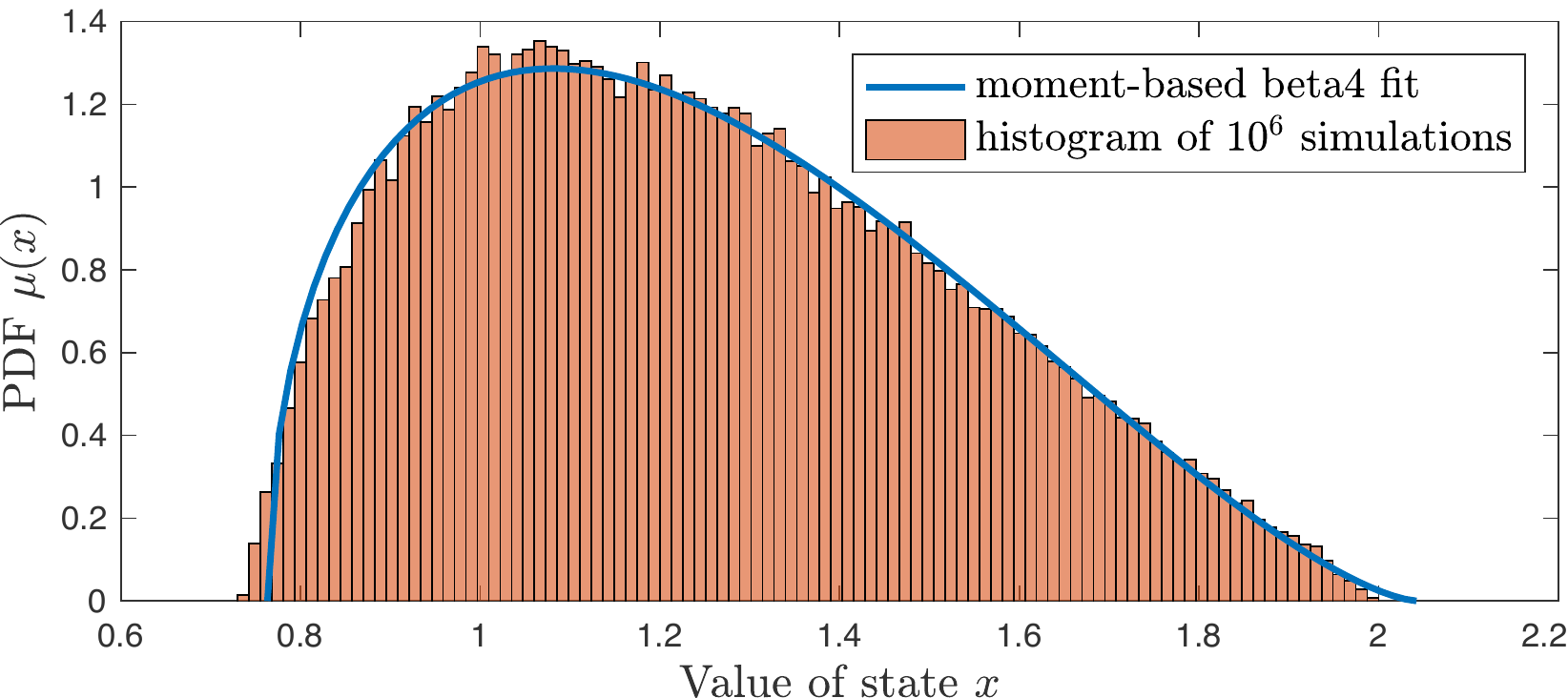}
\caption{PDF $\mu(x)$ after simulating system \eqref{eq:exampleode} for 1\,s and comparison between a  fitted 4-parameter beta distribution and a histogram of $10^6$ samples.}
\label{fig:ex-simple-beta4}
\end{figure}

In order to better visualize the results it might be desirable to fit a probability density functions of the original states, instead of just plotting the moments. 
To do so, PoCET features a routine for recovering a 4-parameter beta distribution from the first four moments, based on \cite{Hanson91}. In the example below, we fist calculate the respective moments for the last time step of the simulation, then recover the shape defining parameters $\alpha$ and $\beta$, as well as the lower and upper bounds of the support.
\begin{lstlisting}
x_beta4 = calcBeta4(x_moments_final);
\end{lstlisting}\vspace{-0.6em}
Fig. \ref{fig:ex-simple-beta4} shows the resulting 4-parameter beta distribution of the state $x$ at the end of the simulation as well as a histogram of respective samples drawn by the Monte-Carlo simulation. Alternatively we could use the PCE coefficients from the system solution to create samples employing the function \textcolor{blue}{\texttt{PoCETsample}}. These can then be used to fit a PDF with one of the available functions in \textsc{Matlab}, like \textcolor{blue}{\texttt{fitdist}} from the Statistics and Machine Learning Toolbox. 

\subsection{Updating the extended system}
\vspace{-0.5em}As long as the distributions of the uncertainies do not change it is not necessary to recompute the entire PCE when their actual values change, since this only corresponds to an update of their respective PCE coefficients. For example, changing $a$ in \eqref{eq:exampleode} from $a\sim\mathcal{B}(2,2)$ to $a\sim\mathcal{B}(4,6)$ can be achieved by calling 
\begin{lstlisting}
pce_sys = PoCETupdate(pce_sys,'a',[4,6]);
\end{lstlisting}\vspace{-0.6em}
where the last input argument is the new \texttt{data} property of the parameter and therefore has to match its expected format (cf. Sec. \ref{sec:sysdef}). 


\section{Advanced examples}\label{sec:examples}
This section provides a general overview of dealing with polynomial and general nonlinear systems as well as an extended example of how PoCET can be used for experimental design. 

\subsection{Higher-order polynomial and nonlinear systems}\label{sec:nonlinear-systems}
\vspace{-0.5em}We can handle polynomial systems directly using a projection-based PCE, since multiplication does not interfere with the orthogonality of the basis polynomials. However, the number of integrals to be solved on setting up the extended system increases significantly. If we consider, for instance, the system
\begin{align}
\dot x &= a(\xi)x^3
\end{align}
we calculate its expansion analogously to \eqref{eq:orth-projection} as
\begin{align}
\begin{split}
\dot{\hat{x}}_i &= \frac{1}{\Braket{\varphi_i,\varphi_i}}\sum_{j,k,l,m=0}^{\tilde{P}-1}a_j\hat{x}_k\hat{x}_l\hat{x}_m\Braket{\varphi_j\varphi_k\varphi_l\varphi_m,\varphi_i}\\
&=\sum_{j=0}^{\tilde{P}-1} \bm{E}^{(3)}_j a_j\bm{\hat x}\otimes\bm{\hat x}\otimes\bm{\hat x}\,.
\end{split}
\end{align}
The dimension of the coefficient matrix is $\bm{E}^{(3)}_j\in\mathbb{R}^{\tilde{P}\times \tilde{P}^3}$, i.e. $\tilde{P}^4$ integrals have to be solved (cf. Eq. \eqref{eq:coeffmat}). 
In order to apply a PCE to a general nonlinear system, it is necessary to use the collocation approach in PoCET, since the involved algorithms to solve the projections in Eq. \eqref{eq:orth-projection} heavily exploit orthogonality of  the polynomials $\varphi_i$, which
does not hold for general nonlinear functions with $\varphi_i$ as arguments. However, most nonlinear systems can be transformed into a polynomial system using immersions, as decribed in \cite{Ohtsuka05}. Note, however, that this approach is only viable for low-dimensional nonlinear systems, since it amounts to an additional state space expansion on top of the one done by the PCE. 

\subsection{Optimal experimental design}
\begin{figure}[b]
\includegraphics[width=\columnwidth]{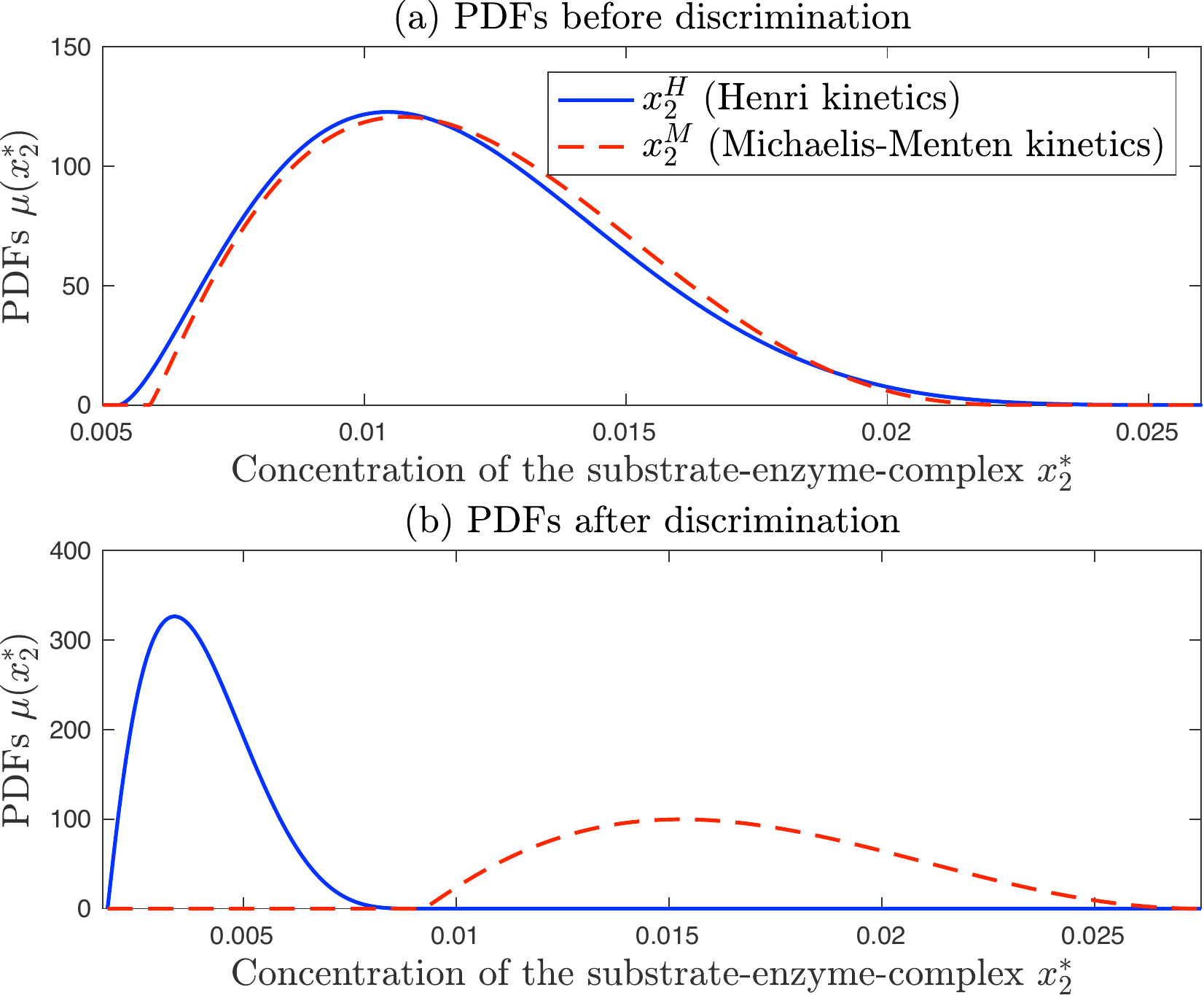}
\caption{Probability density functions of $x_2^*$ after 10\,s without (above) and with (below) discriminating input.} 
\label{fig:oed-pdfs}
\end{figure}

\begin{figure}[b]
\includegraphics[width=\columnwidth]{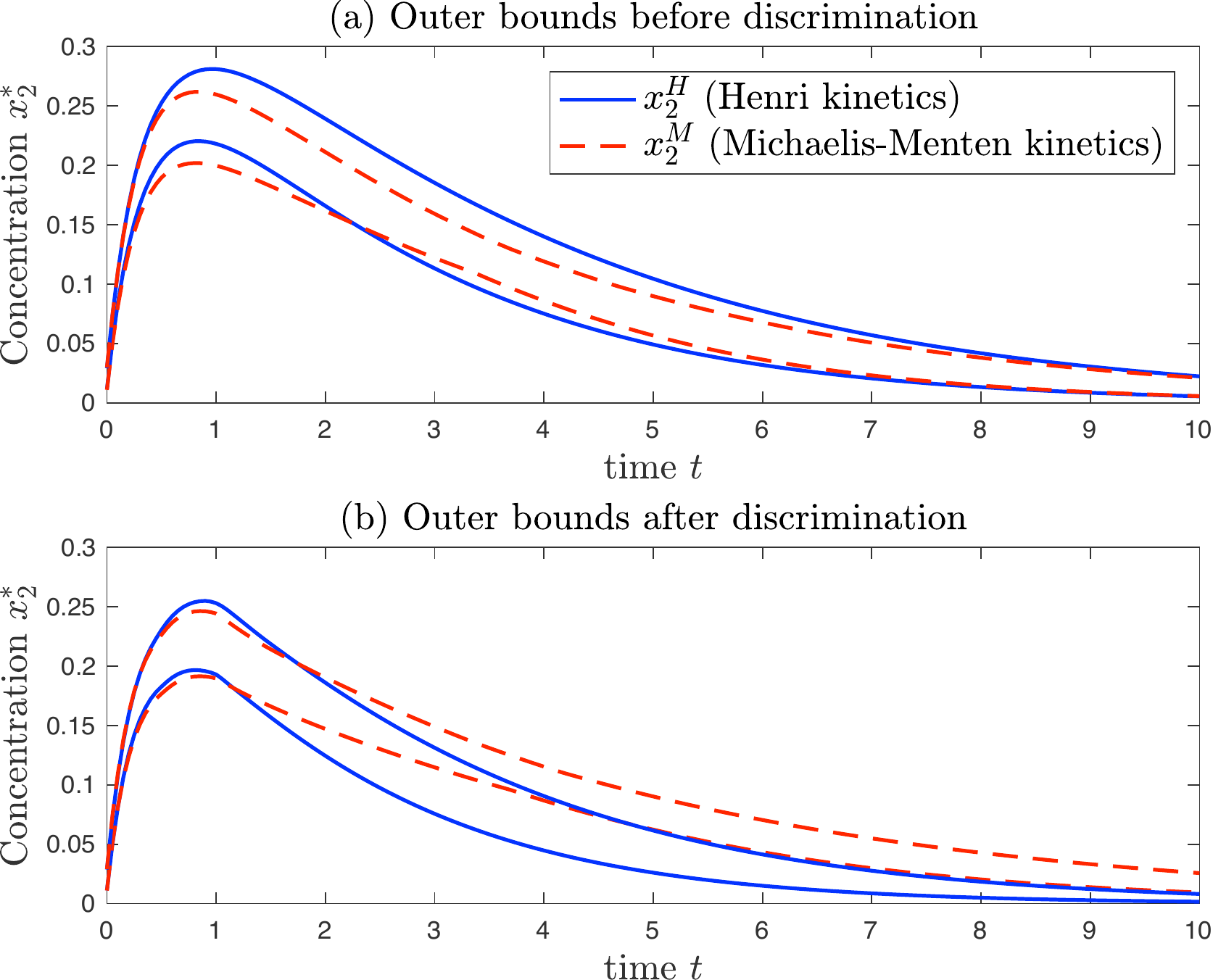}
\caption{Outer bounds of the PDFs of the substrate-enzyme-complex concentrations over time.}
\label{fig:oed-outerbounds}
\end{figure}

\vspace{-0.5em}One of the major advantages of PCE is the low computational effort associated with system simulation compared to sampling-based methods. This makes it highly suitable for tasks that require several simulations like model predictive control problems or optimal experimental design. \\
In the following we consider an optimal experimental design problem previously analyzed in \cite{Henri02, Streif14}, with the Michaelis-Menten and Henri mechanisms as two assumed model hypotheses for an enzyme-catalyzed reaction. 
Both reaction models consider a substrate $S$ and an enzyme $E$, which form an enzyme-substrate complex $C$ and a final product $P$ (cf. \cite{Rumschinski10}). 
The model dynamics follow the law of mass action, where $x_1$ and $x_2$ denote the concentrations of the substrate and complex, respectively. The first model candidate, describing the Henri mechanism, is defined as
\begin{align}
\begin{split}
\dot{x}_1^H&=\left(p_1^H+p_3^H\right)\left(x_2^H-1\right)x_1^H+\left(p_2^H+u\right)x_2^H\\
\dot{x}_2^H&=p_1^H\left(1-x_2^H\right)x_1^H-\left(p_2^H+u\right)x_2^H,
\label{eq:henri}
\end{split}
\end{align}
while the second one, describing the Michaelis-Menten mechanism, is defined as
\begin{align}
\begin{split}
\dot{x}_1^M&=p_1^M\left(x_2^M-1\right)x^M_1+\left(p_2^M+u\right)x_2^M\\
\dot{x}_2^M&=p_1^M\left(1-x_2^M\right)x_1^M-\left(p_3^M+p_2^M+u\right)x_2^M,
\label{eq:micha}
\end{split}
\end{align}
where the input $u$ is assumed to affect the reaction rate $p_2^*$ in an additive manner. Furthermore, we assume uncertainties in the initial concentrations (with $x_1^*(0)\sim\mathcal{B}_4(3, 3, 0.96, 0.98)$ and $x_2^*(0)\sim\mathcal{B}_4(3, 3, 0.01, 0.03)$, respectively) and uncertain estimates for the parameter values (with $p_i^H\sim\mathcal{U}(0.9,1.1)$ and $p_i^M\sim\mathcal{U}(0.9,1.15)$ for all $i=1,2,3$) to be available. Under these uncertainties and assuming that we take a measurement 10\,s after the reaction started, the two models are barely distinguishable with a small number of measurements since the resulting PDFs have a very large overlap, as shown in Figs. \ref{fig:oed-pdfs}(a) and \ref{fig:oed-outerbounds}(a). This makes it almost impossible to decide which model hypothesis is the right one.\\
Our goal is to find a piecewise constant input signal $u$ that discriminates the two models. Defining such an input in PoCET is done in the system definition. While the ODEs for the states are defined as shown in Step \ref{sec:sysdef}. This time, however, we also define an input structure via the code
\begin{lstlisting}
inputs.name = 'u'; 
inputs.rhs  = 'piecewise(u_t,u_v,t)'; % define input as piecewise constant signal
inputs.u_t = [0,1,2,3,4]; % time values at which input value changes
inputs.u_v = [0,0,0,0,0]; % input values to be taken at times defined in u_t
\end{lstlisting}\vspace{-0.6em}
where \texttt{piecewise(u\_t,u\_v,t)} is a function that returns $u_v(k)$ for all times $t\in [u_t(k), u_t(k+1)]$. Note that the additional fields in the input structure follow no predefined scheme -- their names essentially act as additional variables that can be changed in the online simulation. If we want to apply a stairs-shaped input to the system above, we call
\begin{lstlisting}
gal_results = PoCETsimGalerkin(pce_sys,'PCEODE.m',[],simoptions,'u_v',[1,2,3,4,5]); 
\end{lstlisting}\vspace{-0.6em}
where the last input has to match the one defined in the \texttt{inputs} structure above.

Using this, we set up a model-based optimization problem for \texttt{u\_v} employing the \texttt{fmincon} function and the Bhattacharyya distance (cf. \cite{Kailath67}) as a similarity measure for the resulting PDFs. We implemented the latter as a nonlinear constraint in order to enforce the discrimination of the two models.
The optimization took 11.5\,s, including a total of 185 evaluations of the nonlinear constraint (i.e. 370 simulations). Fig. \ref{fig:oed-pdfs}(b) shows the final PDFs under the discriminating input.
Fig. \ref{fig:oed-outerbounds}(b) gives an even stronger result: since it shows the outer bounds of the concentrations $x_2^*$ we can see that any measurement taken after 5\,s or later would allow to decide which model hypothesis is the right one. 
The complete example with additional comments on the usage is included in PoCET, while more details about the employed methods can be found in \cite{Streif14}. 

\section{Further Remarks and Outlook}\label{sec:outlook}
This paper introduced PoCET, a polynomial chaos expansion toolbox for \textsc{Matlab}. Its main contribution is the automated  expansion of polynomial ODE systems, which allows for propagation of probabilistic uncertainties within the projection-based PCE framework. 
PoCET features a very straight-forward system definition and requires little to no knowledge of the PCE framework itself. 
It relies on the Symbolic Math Toolbox for input parsing but is otherwise stand-alone, open-source, and available for free (see \texttt{www.tu-chemnitz.de/etit/control/research/PoCET/} \\for more information). Due to its modular design, it can be adapted into many different directions (e.g. arbitrary PCE by adjusting the quadrature rules for the integration), used for various applications (e.g. stochastic MPC or fault detection), or combined with other uncertainty quantification tools that support polynomial chaos, like UQLab (\cite{uqlab}). 
\bibliography{PCET_IFAC}
\end{document}